\documentclass[rnote,traditabstract]{aa} 
\usepackage{graphicx}
\usepackage{txfonts}
\usepackage{natbib}
\bibpunct{(}{)}{;}{a}{}{,} 
\newcommand{\FWHM}{\mbox{$F\!W\!H\!M$}}
\begin{document}
\title{Multiple emission line components in detached post-common-envelope
binaries%
\thanks{Based on observations made at ESO telescopes (proposals 383.D-0430,
384.D-0596)}}

\author{
C. Tappert\inst{1},
B. T. G\"ansicke\inst{2},
A. Rebassa-Mansergas\inst{1},
L. Schmidtobreick\inst{3},
M. R. Schreiber\inst{1}
}

\authorrunning{C. Tappert et al.}
\titlerunning{Multiple emission components in PCEBs}

\institute{
Departamento de F\'{\i}sica y Astronom\'{\i}a, Facultad de Ciencias,
Universidad de Valpara\'{\i}so, Av. Gran Breta\~na 1111, Valpara\'{\i}so, 
Chile\\
\email{ctappert@dfa.uv.cl, arebassa@dfa.uv.cl, matthias.schreiber@uv.cl}
\and
Department of Physics, University of Warwick, Coventry CV4 7AL, UK\\
\email{Boris.Gaensicke@warwick.ac.uk}
\and
European Southern Observatory, Casilla 19001, Santiago 19, Chile\\
\email{lschmidt@eso.org}
}

\date{Received ; accepted }

\abstract{
Motivated by the recent discovery of an H$\alpha$ emission line component 
originating close to the white dwarf primary in the post-common-envelope binary 
(PCEB) LTT 560, we have undertaken a spectroscopic snapshot survey on 11 
short-period ($P_\mathrm{orb} < 6~\mathrm{h}$) PCEBs using FORS2. We have found 
multi-component H$\alpha$ emission line profiles in six of our targets, 
indicating that multiple H$\alpha$ emission sites are rather common in
short-period PCEBs. The underlying physical mechanisms, however, can be 
fundamentally different. In our sample we find one system where the extracted 
radial velocities of two identified components indicate that both originate 
in the late-type secondary star, where one is caused by irradiation from the 
hot white dwarf primary and the other by stellar activity. Another object
presents two components that are located on opposite sides of the 
centre-of-mass, suggesting a case similar to LTT 560, i.e.~that accretion 
of stellar wind from the secondary star produces H$\alpha$ emission on,
or close to, the white dwarf. 
}


\maketitle
%

\section{Introduction}

The recipe for forming a close interacting binary star such as a cataclysmic
variable contains the common envelope (CE) phase as a crucial ingredient.
Here the more massive star of the detached progenitor binary system 
expands in the course of its nuclear evolution, fills its Roche lobe, and
transfers mass at high rates to its low-mass companion, usually a late-type
(K--M) main-sequence star. The resulting CE provides enhanced friction, thus
causing an accelerated reduction of the binary separation. For details on
the CE phase see \citet{webbink08-1} and \citet{taam+ricker10-1}, and
references therein.

The resulting post-common-envelope binary (PCEB) is a detached white dwarf /
late-type dwarf system with an orbital period of a few days or less.
Its binary separation continues to decrease due to angular momentum loss
driven by magnetic braking and gravitational wave radiation until eventually 
the Roche lobe of the secondary star comes into contact with the stellar 
surface, thus starting mass transfer via Roche lobe overflow.

Since the PCEB is still a detached system one would expect the H$\alpha$
emission line -- if present at all -- to have a simple, single-component
profile originating in the secondary star and produced by stellar
activity or by irradiation of a sufficiently hot white dwarf primary.
However, there are a number of systems known that present a more complex line
profile. For example, in some PCEBs with a hot white dwarf primary, non-LTE
effects produce an inverted core in the irradiation induced emission lines
that give them a double-peaked profile \citep{barmanetal04-1}. As an example
see the recent study of NN Ser by \citet{parsonsetal10-1}. The PCEB QS Vir
shows a complex line profile composed of at least three components that
led \citet{odonoghueetal03-1} to suspect an accretion disc as one of the
origins, while \citet{parsonsetal11-1} suggest stellar prominences as the
mechanism behind it. Finally, the H$\alpha$ line in the system LTT 560 
exhibits two exactly anti-phased components. One component originates in
the secondary star and is related to stellar activity, while the other comes
from a chromosphere/corona on the white dwarf as the probable result of
accretion of stellar wind from the companion star \citep{tappertetal07-1,%
tappertetal11-2}.

Systems of this last type present an attractive possibility to determine a 
full set of parameters (i.e., masses, inclination, binary separation), since 
measuring the gravitational redshift from the H$\alpha$ component on the white 
dwarf yields its mass, and thus breaks the parameter degeneracy of the orbital 
equations. Ultraviolet spectra presented by \citet{kawkaetal08-1} proved that 
ongoing wind accretion in detached PCEBs is a common phenomenon. However, 
LTT 560 so far represents the only known example where the signature of such
accretion is observed as an H$\alpha$ emission line component located on the 
white dwarf. Consequently, it is unclear under what conditions such
chromospheric emission on the white dwarf will develop. One may
assume that it is related to the mass-transfer/accretion rate. PCEBs that are
already comparatively close to contact and/or contain an active secondary star
with a strong stellar wind should thus represent good candidates. The 
probability of fulfilling both, i.e.~having late-type secondary star with 
a large Roche-lobe filling factor, is highest in short-period PCEBs, such
as LTT 560 ($P_\mathrm{orb} = 3.54~\mathrm{h}$). We here present a 
spectroscopic time-series snapshot survey of 11 PCEBs from the Sloan Digital 
Sky Survey \citep[SDSS; ][ and references therein]{abazajianetal09-1} with 
$P_\mathrm{orb} \le 6~\mathrm{h}$.


\section{Observations and data reduction}

\begin{table*}
\caption[]{Log of observations.}
\label{log_tab}
\begin{tabular}{lllllllllll}
\hline\noalign{\smallskip}
(1) & (2) & (3) & (4) & (5) & (6) & (7) & (8) & (9) & (10) & (11) \\
SDSS & date & HJD & $n$ & $t_\mathrm{exp}$ & $\Delta t$ & $\Delta \varphi$ 
& $P_\mathrm{orb}$ & $T_\mathrm{WD}$ & MS & $r$ \\
 & & [d] & & [s] & [h] & orbits 
& [h] & [$10^3$ K] & & [mag] \\
\hline\noalign{\smallskip}
J005245.11$-$005337.2 & 2009-10-17 & 2\,455\,122 &  3 & 900 & 0.52 & 0.19 
& 2.73 & 16 & M4 & 19.1 \\
J015225.38$-$005808.5 & 2009-10-17 & 2\,455\,122 & 10 & 240 & 0.69 & 0.32 
& 2.17 & 9 & M6 & 17.7 \\
J023804.39$-$000545.7 & 2009-10-17 & 2\,455\,122 &  4 & 660 & 0.58 & 0.11 
& 5.08 & 22 & M3 & 18.7 \\
J030308.35+005444.1 & 2009-10-17 & 2\,455\,122 &  7 & 360 & 0.66 & 0.21 
& 3.20 & $-$ & M4 & 18.1 \\
J141134.70+102839.7 & 2009-04-15 & 2\,454\,937 &  3 & 900 & 0.52 & 0.13 
& 4.02 & 30 & M3 & 19.1 \\
J152933.25+002031.2 & 2009-04-15 & 2\,454\,937 &  4 & 660 & 0.58 & 0.15 
& 3.95 & 14 & M5 & 18.3 \\
J155904.62+035623.4 & 2009-04-15 & 2\,454\,937 &  4 & 660 & 0.58 & 0.26 
& 2.27 & 49 & $-$ & 18.6 \\
J211205.31+101427.9 & 2009-06-20 & 2\,455\,003 &  4 & 660 & 0.58 & 0.27 
& 2.17 & 20 & M6 & 18.4 \\
J212320.74+002455.5 & 2009-05-25 & 2\,454\,977 &  6 & 900 & 1.39 & 0.39 
& 3.58 & 13 & M6 & 19.4 \\
J213218.11+003158.8 & 2009-05-26 & 2\,454\,978 &  7 & 360 & 0.66 & 0.12 
& 5.33 & 16 & M4 & 18.2 \\
J221616.59+010205.6 & 2009-05-26 & 2\,454\,978 &  7 & 360 & 0.66 & 0.13 
& 5.05 & 12 & M5 & 18.0 \\
\hline\noalign{\smallskip}
\multicolumn{11}{l}{
\parbox{14.5cm}{
Columns: (1) target's SDSS designation;
(2) start of night of observation;
(3) Heliocentric Julian Day;
(4) number of exposures;
(5) exposure time;
(6) covered time range;
(7) covered fraction of orbital cycle;
(8) orbital period;
(9) approximate temperature of the white dwarf;
(10) spectral type of the secondary star;
(11) magnitude in Sloan $r$.\\
Orbital periods were taken from \citet{rebassa-mansergasetal08-1} and
Nebot Gomez-Moran et al.\, (in prep.), columns (9)--(11) are from 
\citet{rebassa-mansergasetal10-1}.
}}
\end{tabular}
\end{table*}

Spectroscopic time series data were taken with FORS2 
\citep{appenzelleretal98-3} mounted on UT1 (ANTU) at the ESO-Paranal 
Observatory. The instrument was used in long-slit mode (LSS)
with grism 1200R+93 and filter GG435. The covered spectral range was
5750 -- 7230\,{\AA}, and an 0.4'' slit gave a spectral resolution of
2.3\,{\AA}. The observations were conducted in service mode with one
observing block (OB) of less than 1 h per target (2 subsequent OBs for the
faintest one, J2123). Each OB consisted of a series of 3--10 spectra, depending
on the target's brightness and its orbital period. Table \ref{log_tab}
presents a summary of the observations.

Data reduction was performed with IRAF, and included the correction with
bias and flatfield frames. The spectra were extracted, and wavelength 
calibration was obtained with arc lamp spectra that were taken once per night. 
The dispersion solution proved stable over the complete range spanned by the
observations, i.e.\,over several months. The data were not flux calibrated.

Radial velocities of the H$\alpha$ emission line were measured using single
Gaussian functions. In two cases, J1559 and J2216, an additional emission
component was sufficiently visible in at least one spectrum to fit a 
combination of two Gaussians. The resulting full width at half maximum (\FWHM) 
and amplitude of the weaker component were then held fixed to measure the 
radial velocities in the remaining spectra.


\section{Results}

\begin{figure}
\includegraphics[width=\columnwidth]{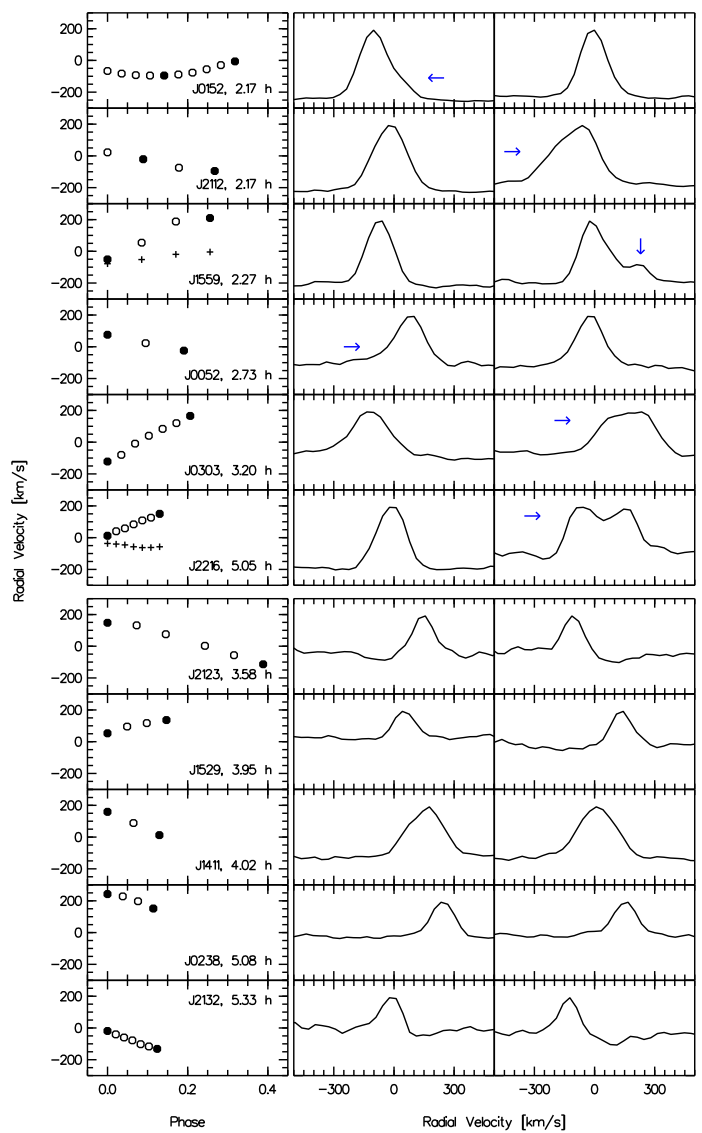}
\caption[]{H$\alpha$ line profiles of the targets, normalised to the intensity
maximum. For each object, two examples profiles are shown that correspond to 
the radial velocities marked as filled circles in the left subpanel. Since
the ephemerides are not known with sufficient precision for an extrapolation,
the zero point of the orbital phase was arbitrarily set to the first data 
point of the respective sets. 
The upper main panel presents the six systems with 
multiple emission sources, with arrows marking the signature of the weaker 
component. The lower panel shows the remaining five systems. The radial
velocities were measured by fitting a single Gaussian to the line profile, 
with the exception of J1559 and J2216, where a combination of two Gaussians 
was used. 
}
\label{lprof_fig}
\end{figure}

\begin{figure}
\includegraphics[angle=-90,width=\columnwidth]{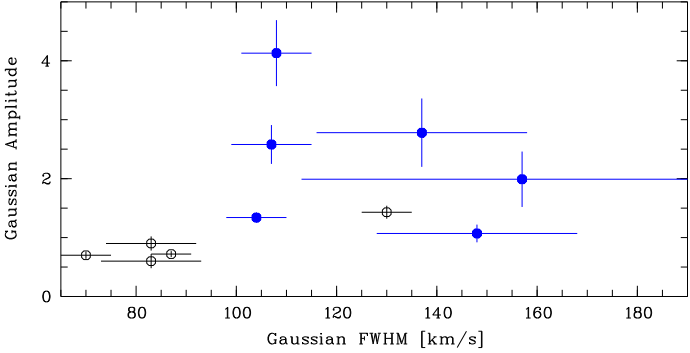}
\caption[]{Average values for \FWHM\, and amplitude of Gaussian fits to the 
line profile for each system. Here, all spectra were fit with a single 
Gaussian, regardless of potentially resolved components. The error bars 
indicate the variability of these parameters over the set of line profiles for 
a specific system. Filled circles mark objects with multiple components.}
\label{fwhm_fig}
\end{figure}

For six of the eleven systems we find a significantly variable emission line 
profile that suggests the presence of at least two emission components
(Fig. \ref{lprof_fig}). Of the remaining five objects, four are very likely
to have a single-component H$\alpha$ emission, while one system, J1411, needs
more data. In the following we review the results for the individual systems.

\smallskip\noindent
{\em J1559, J2216:} In one or more spectra the two components were 
resolved enough to fit individual Gaussian functions to each of
them. For J1559 the two components move together in phase, but with different
amplitudes. The most likely explanation for this behaviour is that both
components originate in the main-sequence star, one due to stellar
activity, the other irradiation. The latter will be more confined to
the white-dwarf facing side, thus be situated closer to the centre-of-mass,
and consequently have a lower radial-velocity amplitude. In J1559, this would
also be the stronger of the two components. We also note that J1559 contains
by far the hottest white dwarf of our sample (Table \ref{log_tab}). With 
$T_\mathrm{WD} = 49\,000~\mathrm{K}$ it has a high probability of an 
irradiation effect (see Section \ref{disc_sec}).

J2216, on the other hand, shows two equally strong components that move in
roughly anti-phased manner. Assuming that one of them is located on the
secondary star, the other then would have to develop on the white-dwarf
side of the centre-of-mass, suggesting a similar situation as in LTT 560.

\smallskip\noindent
{\em J0152, J0303, J2112:} While the individual components are not resolved,
in all three objects we find a strongly asymmetric line profile in the spectra
that are closest to one of the extrema of the radial velocity curve. This is 
clear evidence that the line profile is composed of multiple components.

\smallskip\noindent
{\em J0052:} The asymmetry in the first spectrum of a series of three is
less notable than in above three systems, but it is doubtlessly there. The
radial velocity curve suggests that the three spectra were taken close to
conjunction, when emission components that are in phase or anti-phased would
have very similar velocities.

\smallskip\noindent
{\em J1411:} We do not find any significant line profile variation in this
object. However, as for J0052, our series consists of only three spectra that
were taken close to conjunction as indicated by the radial velocity curve.
Additionally, only 0.13 orbital cycles are covered by the spectra. 

\smallskip\noindent
{\em J0238, J1529, J2123, J2132:} all four systems have weak and narrow
single-component line profiles. These profiles do present some variability, 
but here it is caused by the superposition of the H$\alpha$ emission
from the secondary with the broader H$\alpha$ absorption of the white dwarf
primary that move in anti-phase. 

\smallskip\noindent
In Fig.~\ref{fwhm_fig} we show the average \FWHM\, of a Gaussian function fit 
to the series of line profiles of a specific object versus its amplitude. The 
assumption is that lines composed of more than one component have, on average, 
a broader profile and, as the multiple components merge and separate with 
orbital phase, show a larger variation in both intensity and width when fitted 
with a single Gaussian than genuine single-component lines. While aware 
of looking at small number statistics, we do note that the two types of systems 
are clearly separated, with all multiple-component objects having 
$\FWHM_\mathrm{av} > 100~\mathrm{km/s}$, while $\FWHM_\mathrm{av} < 
90~\mathrm{km/s}$ for all single-component systems, with J1411 
($\FWHM_\mathrm{av} = 130~\mathrm{km/s}$) being the sole exception. 
Considering the width of the line profile and the small orbital phase coverage 
that additionally is likely to be close to conjunction, we thus conclude that 
with the present data we cannot exclude with certainty the presence of 
multiple emission components in J1411.
In this context we note that our spectral resolution of $\sim$2.3 {\AA}
translates to $\sim$100 km/s at H$\alpha$. This explains the clustering
of the single-component systems with $\FWHM_\mathrm{av} < 90~\mathrm{km/s}$,
which thus is simply due to our limited spectral resolution. Consequently the
respective line profiles are not resolved, and the real $\FWHM$ is
probably even lower for those systems.

\section{Discussion\label{disc_sec}}

\begin{table*}
\caption[]{Comparison of WD/MS PCEBs with $P_\mathrm{orb} < 1~\mathrm{d}$ that
have time-series spectroscopic data published at sufficient high spectral
resolution to examine the H$\alpha$ emission line profile for the presence
of multiple components.}
\label{pceb_tab}
\begin{tabular}{lllllll}
\hline\noalign{\smallskip}
(1) & (2) & (3) & (4) & (5) & (6) & (7) \\
object & $P_\mathrm{orb}$ & $T_\mathrm{WD}$ & MS 
& $n_{\mathrm{H}\alpha}$ & H$\alpha_\mathrm{MS}$ & references \\
 & [h] & [$10^3$ K] & & & & \\
\hline\noalign{\smallskip}
J0152        & 2.17  & 9  & M6   & 2   & --  & this paper \\
J2112        & 2.17  & 20 & M6   & 2   & --  & this paper \\
J1559        & 2.27  & 49 & --   & 2   & a+i & this paper \\
HR Cam       & 2.47  & 19 & --   & (1) & i   & \citet{bergeronetal92-3,%
marsh+duck96-1,maxtedetal98-1}\\
J0052        & 2.73  & 16 & M4   & 2   & --  & this paper \\
HS 2237+8154 & 2.98  & 12 & M3.5 & 2   & a   & \citet{gaensickeetal04-2} \\
NN Ser       & 3.12  & 57 & M4   & (1) & i   & \citet{haefner89-1,%
haefneretal04-1,parsonsetal10-1} \\
J0303        & 3.20  & -- & M4   & 2   & --  & this paper \\
LTT 560      & 3.54  & 7  & M5.5 & 2   & a   & \citet{tappertetal07-1,%
tappertetal11-2} \\
J2123        & 3.58  & 13 & M6   & 1   & --  & this paper \\
QS Vir       & 3.62  & 14 & M4   & 3   & a   &  \citet{kawkaetal02-1,%
odonoghueetal03-1,parsonsetal11-1}\\
J1529        & 3.95  & 14 & M5   & 1   & --  & this paper \\
J1411        & 4.02  & 30 & M3   & 1?  & --  & this paper \\
BPM 71214    & 4.85  & 17 & M2.5 & 1   & a   & \citet{kawkaetal02-1,%
kawka+vennes03-1} \\
J2216        & 5.05  & 12 & M5   & 2   & --  & this paper \\
J0238        & 5.08  & 22 & M3   & 1   & --  & this paper \\
J2132        & 5.33  & 16 & M4   & 1   & --  & this paper \\
HS 1857+5144 & 6.38  & 85 & M5   & (1) & i   & \citet{aungwerojwitetal07-1}\\
CC Cet       & 6.89  & 26 & M5   & 1   & i   & \citet{safferetal93-1,%
somersetal96-3} \\
RR Cae       & 7.30  & 8  & M4   & 1   & a   & \citet{bruch+diaz98-1,%
bruch99-1,maxtedetal07-1} \\
WD 1042-690  & 8.06  & 21 & --   & 1   & a   & \citet{bragagliaetal95-1,%
morales-ruedaetal05-1} \\
DE CVn       & 8.74  & 8  & M3   & 1   & a   & 
\citet{vandenbesselaaretal07-1} \\
V471 Tau     & 12.50 & 35 & K2   & 1   & --  & \citet{nelson+young70-1,%
young+nelson72-1};\\
& & & & & & \citet{gilmozzi+murdin83-1,boisetal91-1} \\
HZ 9         & 13.54 & 17 & M4.5 & 1   & a   & \citet{lanning+pesch81-1,%
bleachetal02-1,schreiber+gaensicke03-1}\\ 
UZ Sex       & 14.33 & 20 & M4   & 1   & a   & \citet{safferetal93-1,%
kepler+nelan93-1,bruch+diaz99-1} \\
EG UMa       & 16.03 & 13 & M4   & 1   & a   & \citet{bleachetal00-1,%
bleachetal02-1} \\
WD 2009+622  & 17.78 & 26 & M5.5 & 1   & i   & \citet{bergeronetal92-3,%
morales-ruedaetal05-1,farihietal05-7} \\
HS 1136+6646 & 20.06 & 70 & K5.5 & (1) & i   & \citet{singetal04-1} \\
\hline\noalign{\smallskip}
\multicolumn{7}{l}{
\parbox{18cm}{Columns: (2) orbital period, (3) approximate temperature of
the white dwarf, (4) spectral type of the main-sequence star, (5) number of
H$\alpha$ emission components, a `1' in brackets marks non-LTE effects,
(6) origin of the H$\alpha$ component from the main-sequence star, either
stellar (a)ctivity or (i)rradiation from the white dwarf.
}}
\end{tabular}
\end{table*}

\begin{figure}
\includegraphics[width=\columnwidth]{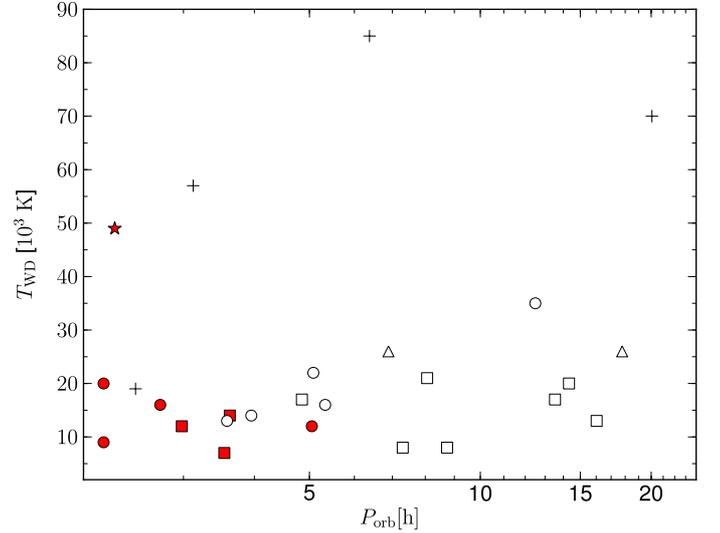}
\caption[]{Orbital period (in logarithmic scale) vs.~white-dwarf temperature
for the systems in Table \ref{pceb_tab}. Filled symbols mark PCEBs with 
multiple H$\alpha$ emission components, open symbols those with single
components. The shape of the symbol indicates the type of emission from the
secondary star: irradiation (triangle), activity (square), or yet unknown
(circle). J1559 is marked by a star, since it probably presents both
effects. Objects that show non-LTE effects are indicated by crosses. 
The plot does not include the systems J0303 ($T_\mathrm{WD}$ unknown) and 
J1411 (uncertain number of emission components).
}
\label{porbtwd_fig}
\end{figure}

In a snapshot survey of 11 short-period PCEBs we have found strong evidence
of multiple H$\alpha$ emission sources in six systems. Due to the limited
phase coverage of our spectroscopy it is not possible to determine the 
location within the binary system of any of those sources. One might expect 
that at least one of them is located on the secondary star, with the 
underlying mechanism either stellar activity or irradiation by the white 
dwarf primary. However, in J1559 we find two components that move in phase but 
with different amplitudes, which can be interpreted as both these effects 
being present at the same time. In contrast, J2216, the only other system 
where we can actually measure the velocities of the individual components, 
presents two approximately anti-phased emission components, which then should 
come from opposite sides of the centre-of-mass. 

These two examples already highlight that very different physical mechanism 
are underlying the multiple H$\alpha$ components in PCEBs. So far, we can 
discern two fundamentally different scenarios for emission line profiles that 
deviate from those in single late-type dwarfs. On the one hand, irradiation 
from a hot white dwarf leads to non-LTE effects \citep[such as in NN Ser; ][]%
{parsonsetal10-1} or to a two-component profile that combines the 
recombination emission with the one from stellar activity, such as in J1559. 
On the other hand, mass transfer can lead to prominences \citep[QS Vir; ][]%
{parsonsetal11-1} or a white-dwarf corona \citep[LTT 560; ][]{tappertetal11-2}.

Despite the complex phenomenology one may ask if it is possible to find a 
common property that would at least allow to discern those PCEBs with multiple 
emission sources from those with single components. To explore the 
parameter space we enlarged the sample by taking the PCEB lists of 
\citet{morales-ruedaetal05-1} and \citet{shimanskyetal06-1}, examined the 
published data on these systems, and included HS 1857+5144 
\citep{aungwerojwitetal07-1}. Since our main interest concerns potential 
precataclysmic variables such as LTT 560, we limited the sample to PCEBs with 
$P_\mathrm{orb} < 1~\mathrm{d}$ that contain a white dwarf as primary 
component. 

The systems with data that allow an examination of the H$\alpha$ 
emission line profile are presented in Table \ref{pceb_tab}. Apart from the 
six PCEBs with multiple emission components from the present survey and the 
known objects LTT 560 and QS Vir, we find one other such system 
\citep[HS 2237+8154; ][]{gaensickeetal04-2}. While it is not explicitly 
mentioned by the authors, the trailed spectrum of the H$\alpha$ emission line 
(their Fig.~8) clearly shows the presence of both a high-amplitude and a 
low-amplitude component that are roughly anti-phased. For the system
HS 1136+6646 \citet{singetal04-1} mention an accretion disc as the possible
mechanism behind the double-peaked H$\alpha$ profile, but the extremely
hot white dwarf primary instead suggests a non-LTE effect.

Table \ref{pceb_tab} includes parameters that might have an influence on the 
number of emission sources. It is clear that these have to be such that they 
facilitate the interaction between the two stellar components. A short orbital 
period basically equals a small binary separation (with the limited mass ranges 
involved here), and a high white-dwarf temperature increases the probability 
of an irradiation effect or even a non-LTE phenomenon%
\footnote{A notable exception is the system HR Cam, which shows non-LTE 
features \citep{maxtedetal98-1}, but does not contain a particularly hot white 
dwarf \citep[$T_\mathrm{WD} \sim 19\,000~\mathrm{K}$; ][]{bergeronetal92-3}}. 
Additionally the spectral type of the secondary star is broadly
correlated with the strength of a stellar-activity induced H$\alpha$ emission 
line \citep[e.g., ][]{berger06-1}, although with the very limited range of 
spectral types in our sample, this should not be a major factor. Other 
potentially important parameters are the magnetic field strength of the white 
dwarf and the Roche-lobe filling factor of the secondary star. While there is 
unfortunately no such data available for most systems, we note that there 
is at least no positive detection of a strongly 
magnetic white dwarf in any of the objects in Table \ref{pceb_tab}. Also, 
while for PCEBs that eventually evolve into cataclysmic variables one expects 
to find nearly Roche-lobe filling secondaries anywhere in an orbital period 
range roughly from 1.5 to 10 h, the probability of a large Roche-lobe filling 
factor will be higher for shorter orbital periods.  

We can see in Fig.~\ref{porbtwd_fig} that these basic assumptions are 
confirmed. PCEBs with multiple emission components tend to cluster at
short orbital periods, and those with single components are mostly confined
to long orbital periods, $> 4.5~\mathrm{h}$. In fact, with the exception of
non-LTE objects we do not find single-component PCEBs below $P_\mathrm{orb}
= 3.5~\mathrm{h}$. Furthermore, irradiation effects dominate at $T_\mathrm{WD}
> 25\,000~\mathrm{K}$ and do not appear to play any role for $T_\mathrm{WD}
< 20\,000~\mathrm{K}$. However, these limits have to be taken with caution
since with the available data the danger of overinterpreting small number 
statistics is all too present. 

For a significant fraction of the systems in our sample the origin of the 
emission from the secondary star is not known, and for the vast
majority of the PCEBs with multiple emission sources we still do not
have any information on the mechanism behind that phenomenon. In fact, so 
far the origin of both components has been firmly established only for LTT 560 
\citep{tappertetal11-2}. The present work should therefore encourage more 
detailed studies, since it is clear that the presence of multiple emission 
components in PCEBs is not an isolated, but rather a common phenomenon. 
On the basis of the currently present data we can conclude that 
the best place to look for it are short-period PCEBs ($P_\mathrm{orb} 
< 4~\mathrm{h}$) with a white dwarf primary that is not too hot ($T_\mathrm{WD} 
< 20\,000~\mathrm{K}$).

\begin{acknowledgements}
We thank the referee, John Thorstensen, for helpful comments.
ARM acknowledges financial support from Fondecyt in the form of grant
number 3110049.
This study has made extensive use of NASA's Astrophysics Data System 
Bibliographic Services. IRAF is distributed by the National Optical Astronomy 
Observatories.
\end{acknowledgements}


\end{document}